\documentclass{article}
\usepackage{amsmath}
\usepackage{amsfonts}
\usepackage{amssymb}
\usepackage{siunitx}
\usepackage{stmaryrd}
\usepackage{bm}
\usepackage[utf8]{inputenc}
\usepackage{gensymb}
\usepackage{lipsum}

\usepackage{graphicx}
\usepackage{tikz}
\usepackage{listings}
\usepackage{pgfplots}
\usepackage{algorithm}
\usepackage{algorithmic}
\usepackage{multicol}
\usepackage{multirow}
\usepackage{enumitem}
\usepackage{mwe}    
\usepackage{caption}
\usepackage{subcaption}
\usepackage{hyperref}
\usepackage{lineno}
\usepackage{authblk}

\usepackage[numbers]{natbib}
\providecommand{\keywords}[1]
{
  \small	
  \textbf{\textit{Keywords---}} #1
}

\title{Contact between rough rock surfaces using a dual mortar method}
\author[1]{Cyrill von Planta}
\author[2]{Daniel Vogler}
\author[1]{Patrick Zulian}
\author[2]{Martin O. Saar}
\author[1]{Rolf Krause}

\affil[1]{Universit\`{a} della Svizzera Italiana, Institute of Computational Science, Lugano, Switzerland}
\affil[2]{ETH Zurich, Geothermal Energy and Geofluids Group, Institute of Geophysics, Zurich, Switzerland}

\begin{document}
\maketitle
\begin{abstract}
The mechanical behavior of fractures in solids, such as rocks, has strong implications for reservoir engineering applications. Deformations, and the corresponding change in solid contact area and aperture field, impact rock fracture stiffness and permeability thus altering the reservoir properties significantly. Simulating contact between fractures is numerically difficult as the non-penetration constraints lead to a nonlinear problem and the surface meshes of the solid bodies on the opposing fracture sides may be non-matching. Furthermore, the challenging geometry of the arising constraints requires to solve the problem in several iterations, adjusting the constraints in each one. Here we present a novel discrete implementation of a dual mortar method and a non-smooth SQP solver, suitable for parallel computing, and apply it to a two body contact problem consisting of realistic rock fracture geometries from the Grimsel underground laboratory in Switzerland. 
The contributions of this article are: 1) a novel, parallel implementation of a dual mortar method and non-smooth SQP method, 2) realistic rock geometries with rough surfaces, and 3) numerical examples, which prove that the dual mortar method is capable of replicating the nonlinear closure behavior of fractures, observed in laboratory experiments.
\end{abstract}

\section{Introduction}
Mechanical contact of rough surfaces of fractures or faults along discontinuities in the rock mass is of significant importance in reservoir engineering applications, such as enhanced geothermal systems (EGS) or CO$_2$ sequestration \citep{AGE+18, tester_2006}.
The mechanical contact determines the normal and the shear compliances of rock fractures, which govern the geomechanical reservoir behavior and thereby influence economical efficiency and risk (e.g., induced seismic hazards or CO$_2$ leakage) of the reservoir application in question \cite{mcclure_2014b}.
The mechanical behavior of rock fractures has therefore been studied extensively, with many studies focusing on surface topography \cite{bandis_1983,zangerl_2008,vogler_2017} and the influence of surface topography on the mechanical behavior of the fracture \cite{bandis_1983,pyraknolte_2000,jiang_2006,matsuki_2008,tatone_2015,vogler_2018}.

Specifically, surface topographies determine the contact area distribution of fractures for given loading conditions, and are thus crucial, when investigating fracture strengths in normal or shear directions.
As a fracture is exposed to mechanical loading normal to the fracture, or loading is increased, fracture closure and load exhibit a nonlinear relationship, with fracture closure increments becoming smaller with increasing load increments and displaying convergent behavior \cite{bandis_1983,matsuki_2008,zangerl_2008,vogler_2016,vogler_2018,kling_2018}. This nonlinear closure behavior of fractures can be attributed to the increase in contact area with fracture closure, which increases the fracture stiffness.

The described rock fracture surface topographies are complex, with characteristic features on the sub-millimeter scale \cite{vogler_2017}. Therefore, computing the contact and deformation of two such fracture surfaces constitute a two- or multi-body contact problem, a problem class that even in the standard finite element (FEM) setting with linearized contact, still poses a significant computational challenge to this day.
This is the case for two reasons: First, the formulation of the non-penetration condition itself is nontrivial as the bodies in consideration normally have non-matching triangulations at the contact boundary. Secondly, the non-penetration condition at the a priori unknown contact zone between the bodies constrains the solution space of possible displacements, thereby introducing a nonlinearity.
Simulation tools that resolve small-scale roughnesses, while allowing computation of contact stresses and stress variations in the rock mass, are rare. Most analytical or numerical studies at the laboratory or reservoir scale rely on simplifying assumptions and regularizations, such as averaging asperity scale processes with an empirical relationship, reduction to two dimensions, using variants of boundary element methods, not modeling explicitly the mechanical contact between two surfaces, treating the surfaces as two parallel plates, using penalty methods or only matching triangulations at the surface \cite{bandis_1983,nemoto_2009,cappa_2011,min_2013a,derode_2013,figueiredo_2015b, SNS+15,RMP12,Mao05, CPH17}. 

 
In this work, we present a new parallel implementation of an approach that does not require any of the previously mentioned assumptions. The approach is conventional in as much as we employ linear elasticity and linearized contact conditions in the strong formulation \cite{KO88}. However, when discretizing with finite elements, we use non-uniform meshes with non-matching surface meshes at the contact boundary and then employ a mortar projection to resolve the contact conditions across the non-matching contact surfaces. We follow \cite{WK03} in that we use a mortar projection with dual Lagrange multipliers and apply a change of basis transform to the system, to effectively transform the two-body problem into a one-body problem. The formulation and discretization rest on a solid theoretical basis with proven numerical stability (see \cite{BHL98, Bel99, Woh00, WK03} and references cited therein) and can be extended to also include friction \cite{Kra09} and multigrid acceleration for the solution of the constrained system \cite{WK03,DK09_mb,Kra09}.
For rough rock surfaces, our mortar technique is particularly useful, since  we have complex, non-matching triangulations at the contact boundary (see Figure \ref{fig:mesh}) and an unknown contact zone. Still, we want to obtain accurate three-dimensional representations of a fracture with contact areas to model fluid flow in fractures. For this, penalty-based approaches are at  a disadvantage, as the solution can be distorted by over-closure or additional convergence problems result from large penalty parameters. In the mortar case, the non penetration condition is enforced locally in a weak sense. Thus, while locally some interpenetration is possible, the discretization error depends on the local mesh size and can, if needed, be reduced at will through local refinement.

The use of this mortar technique has previously been hindered by the absence of an efficient way to compute the projection operator. The assembly of the operator requires the detection of intersections between the non-matching surface meshes. This is a complex task in parallel, as the portions of the intersecting meshes might reside on different compute nodes. To this end we use the relatively recently introduced MOONolith library \cite{KZ16} together with libMesh \cite{KPH+06}.  This, and the reliance on a finite element formulation, makes it feasible to apply the method to a wide range of problems, since the method can be incorporated into the vast number of available finite element software packages. Here, we combine the projection operator with custom components from the MOOSE framework \cite{GNH+09} and PETSc \cite{BGM+97,BAA+17} to conduct our numerical experiments. 

So far, only a few studies have applied dual mortar approaches for geophysics problems \cite{TZW+18, WZZ+19}, the focus being on macro scale problems with smooth surfaces.  This article on the other hand focuses on the contact between rough surfaces at the micro scale, with non-vanishing initial gaps. To ensure a thorough introduction, we will emphasize the fundamentals of the dual mortar method, and not focus on the specifics of the implementation, which is available online.

This paper is organized as follows. In Section \ref{sec:methods}, we formulate the two-body contact problem.  Then, we introduce the mortar projection and its discrete assembly, followed by the change of basis transformations and solution method to solve the resulting system.  In Section~\ref{sec:results}, we show the characteristic boundary stresses from Hertzian contact and simulations of rough fractures, subjected to increased normal loads, which exhibit the characteristic nonlinear closing behavior of a fracture under increased normal stress.
\section{Methods} \label{sec:methods}
\begin{figure}[htbp]
\begin{center}
{\small
 \includegraphics[width=0.8\textwidth]{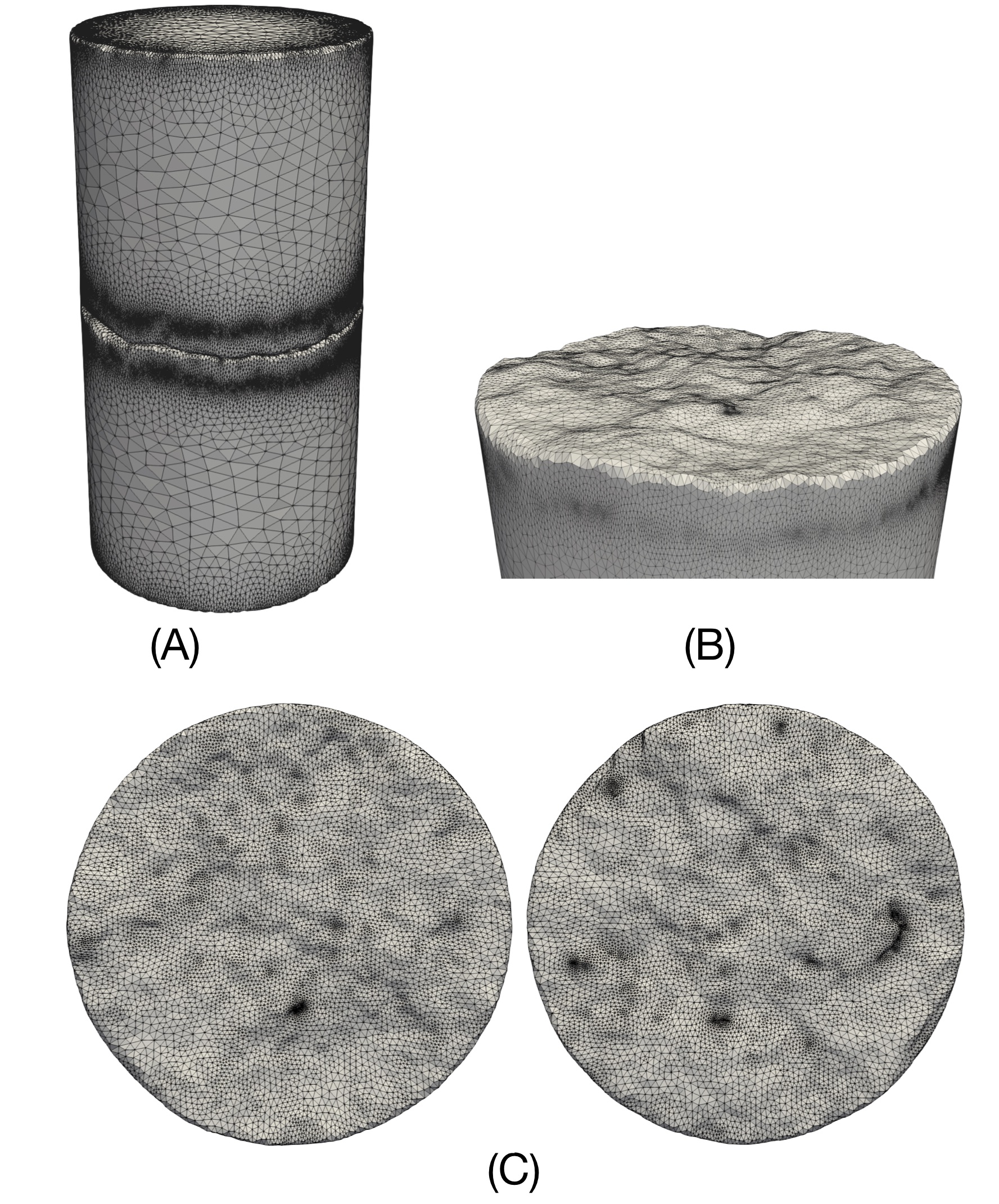}
}
\caption{Mesh used for the numerical simulations with: A) The two specimen halves, which are separated by the fracture; B) The rough fracture surface of the lower specimen half; C) Normal view of the two non-matching meshes of the fracture surfaces.}
\label{fig:mesh}
\end{center}
\end{figure}

In this section we introduce the dual mortar approach and the basis transformations needed to solve the two-body contact problem. An  example for two-body contact on non matching grids is shown in Figure~\ref{fig:mesh}, where the mesh discretization on the fracture surface is shown for a cylindrical specimen with a fracture normal to the cylinder axis in the center.

The dual mortar approach for multibody contact problems is by now a thoroughly studied subject (see \cite{Woh00,WK03,DK09_mb} and references cited therein). However, we believe that these formulations are still not well known in the geophysics community and we will therefore lay a strong emphasis on the formulation of the contact conditions, the dual mortar approach and the change of basis of the overall system. In particular, we would like to point the readers' attention to the successive metamorphosis of the non-penetration condition from a strong (Eq.~\ref{eq:nonpen}) to a weak (Eq.~\ref{eq:weak-nonpen}) and, finally, using the mortar operator, to a discrete form (Eq.~\ref{eq:discrete-nonpen}).

\subsection{Contact formulation} \label{sec:contact}
For the formulation of the contact problem, we try to be as close as possible to the  formulation and notation in \cite{WK03,DK09_mb} and we refer the reader to these articles, and the references cited therein, for a more in-depth introduction.

\begin{figure}[bt]
\begin{center}
{\small
\includegraphics[width=0.9\textwidth]{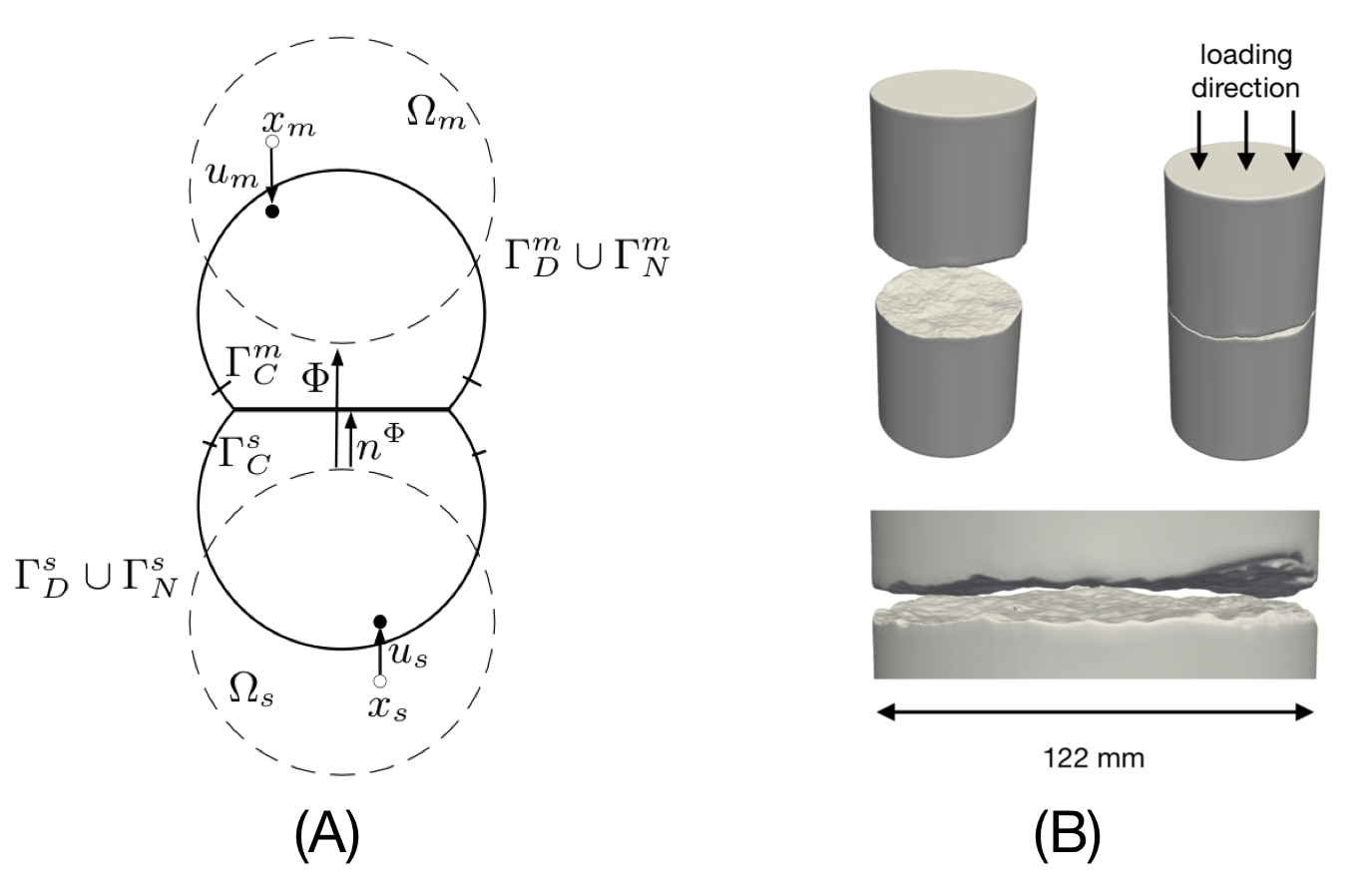}
}
\caption{Two-body problem: A) Schematic of the contact problem between two spheres. The dotted lines indicate the initial position of the bodies, the solid lines show the contact configuration. B) Two-body problem for rough rock fractures. Two rock cylinders in open configuration (left), in closed configuration (right), and a close-up of the rough contact zone (bottom).} 
\label{fig:2bodyproblem}
\end{center}
\end{figure}

We consider a master body $\Omega^m \subset \mathbb{R}^3$ and a slave body $\Omega^s \subset \mathbb{R}^3$. The choice, which body takes which role is arbitrary. We use the superscript $\alpha \in \{m,s\}$ for definitions that apply for both the master and the slave body. For example, the boundary $\Gamma^{\alpha},\, \alpha \in \{m,s\}$, of each body consists of three non overlapping parts: of a Neumann boundary $\Gamma_N^{\alpha}$, a Dirichlet boundary $\Gamma_D^{\alpha}$ and a boundary $\Gamma_C^{\alpha}$, where the possible contact occurs.  Conversely we omit it, when we denote the union of master and slave body, i.e. $\Omega = \Omega^m \cup \Omega^s$ and so on. The displacement field on the bodies $u:= [u^m, u^s]$ is separated into displacements on the master and slave body, respectively (see also Fig. \ref{fig:2bodyproblem}).  The material of $\Omega^{\alpha}$ is considered to be linear elastic.  By enforcing the summation convention on repeating indices from 1 to 3, Hooke's tensor $(E^{\alpha}_{ijml})^3,\,  1 \leq i,j,l,m \leq 3 $ is used to formulate the stresses $\sigma$ given by Hooke's law, using the index $,_j$  to abbreviate derivatives with respect to $x_j$:
\begin{equation} 
 \sigma_{ij}(u^{\alpha}) = E^{\alpha}_{ijml} u^{\alpha}_{l,m}.
\end{equation}
We assume that a bijective mapping $\Phi: \Gamma_C^s \rightarrow \Gamma_C^m$ exists, which maps the points on the slave side of the boundary to the possible contact point on the master side. We then define the vector field of normal directions $n^{\Phi}$:
\begin{equation}
n^{\Phi} : \Gamma^s_C \rightarrow \mathbb{S}^2, \qquad n^{\Phi}(x)   := \left\{
                \begin{array}{lll}
                  \frac{\Phi(x)-x}{|\Phi{x}-x|}  &\quad & \text{if } \Phi(x) \neq x \, \text{(no contact)} \\
                  n^s(x) &\quad & \text{otherwise}
                \end{array}
              \right .
\end{equation}
The gap function $g: \mathbb{R}^3 \rightarrow \mathbb{R} \label{eq:g}, \,  x \mapsto | \Phi(x) -x |$  then measures the width of the gap between the two bodies in the normal direction (i.e., aperture of the fracture in geophysics) and we also define the point-wise jump $[u]  :=  (u^s  - u^m \circ \Phi)\cdot n^{\Phi} $, which is to be smaller than the gap $g$, i.e. $[u]  \leq  g$. This condition is only meaningful in the linearized contact setting, where the bodies are close together and the outer normals $n^{\alpha}\, , \alpha \in \{m, s\}$ are parallel, i.e. we have $n^s := n^{\Phi}$ and $n^m := - n^s$.

For the contact conditions we need stresses and displacements with respect to the outer normal direction and the tangential direction $t$: 
\begin{equation}
\begin{array}{cc}
\sigma_n^{\alpha} = n_i^{\alpha} \cdot \sigma_{ij}(u^{\alpha}) \cdot n_j^{\alpha}, \quad & u_n^{\alpha} = u^{\alpha} \cdot n^{\alpha},\\
\sigma_t^{\alpha} = \sigma(u^{\alpha}) \cdot n^{\alpha} - \sigma_n\cdot n^{\alpha}, \quad & u_t^{\alpha} = u^{\alpha} - u_n^{\alpha} \cdot n^{\alpha}.
\end{array}
\end{equation}

With this we state the contact problem in its strong form. We assume that the body is in an equilibrium state with body forces $f$ (Eq. \ref{eq:equilibrum}), Dirichlet boundary conditions (Eq. \ref{eq:dc_bnd_cnd}), Neumann boundary conditions with pressure $p_i$ (Eq. \ref{eq:nm_bnd_cnd}) and contact boundary conditions (Eq. \ref{eq:con_bnd_cnd}).  As contact conditions we have the non-penetration condition \ref{eq:nonpen}, the complementary condition \ref{eq:compl}, and lastly Equation \ref{eq:nofriction} which states that we have no stresses in tangential directions, i.e., we are considering frictionless contact:
\begin{eqnarray}
-{\rm div}\,\sigma(u) &=& f \quad \mbox{in} \quad  \Omega^s \cup \Omega^m \label{eq:equilibrum}  \\
u_i &=& 0 \quad   \mbox{on} \quad \Gamma_D^{\alpha}  \label{eq:dc_bnd_cnd}\\
\sigma_{ij}(u) \cdot n_j &=&  p_i  \quad  \mbox{on} \quad \Gamma_N^{\alpha}    \label{eq:nm_bnd_cnd} \\
\sigma_n &\leq&  0 \quad   \mbox{on} \quad \Gamma_C. \label{eq:con_bnd_cnd} \\
	\sigma_n (u^m \circ \Phi)	&=&  \sigma_n (u^s) \quad \mbox{on} \quad \Gamma^s_C \\
  	\left[u \right]  &\leq& g \quad \mbox{on} \quad \Gamma^s_C  \label{eq:nonpen}\\
  	\big( [u ] - g \big) \sigma_n(u^s) &=& 0 \quad \mbox{on} \quad \Gamma^s_C  \label{eq:compl}\\
	\sigma_T &=& 0 \quad \mbox{on} \quad \Gamma_C \label{eq:nofriction}.
\end{eqnarray}
For the weak formulation we use the space $X:= X(\Omega^{m}) \times X(\Omega^s)$,  where $X(\Omega^{\alpha}) = [\textit{H}^1_0(\Omega^{\alpha})]^3, \alpha \in \{m,s\}$ are the usual Sobolev spaces, satisfying the Dirichlet boundary conditions (Eq. \ref{eq:dc_bnd_cnd}) and we define the bilinear form $a(u,v)$:
\begin{equation}
a(u,v):= \sum_{\alpha \in {m,s}} \int_{\Omega^{\alpha}} E_{ijlm}^{\alpha} u_{k,j} v_{l,m} \, dx \qquad w,v \in X,
\label{eq:linel_bilinear}
\end{equation}
and the linear functional $\mathbb{f}(v) := \int_{\Omega} f v \, d \omega  + \int_{\Gamma_n} p v \, d\gamma$. Furthermore, we introduce the convex set of admissible displacements $\mathcal{K} \subset X$:
\begin{equation}
\mathcal{K} := \{  u \in X | [u]  \leq g \, ,\text{a.e. on } \Gamma_C^s\}.
\end{equation}
Here the inequality needs to be interpreted pointwise. We can now state the contact problem in its weak form as the minimum of the energy functional $J(u):= \frac{1}{2}a(u,u) - f(u)$:  Find a $u \in \mathcal{K}$ such that:
\begin{equation}
J(u) \leq J(v),  \qquad \forall v \in \mathcal{K}.
\end{equation}
We end this subsection by introducing the finite element spaces $X_h := X_h(\Omega^m) \times  X_h(\Omega^s)$ associated to $X$ and appropriate triangulations $\mathcal{T}_h^m$ and $\mathcal{T}_h^s$ of $\Omega^m$ and $\Omega^s$ with mesh width $h$. Using the basis functions $\phi_i$ of $X_h$ we define 
\begin{equation}
\mathbf{A} :=  a(\phi_i e_k,\phi_j e_l) \text{ and } \mathbf{f}  :=  \mathbb{f}(\phi_i e_k), \quad i,j=1,...,N;k,l=1,...,3,
\label{eq:discrete_operator}
\end{equation}
where $N$ is the number of nodes of the meshes  $\mathcal{T}_h^{\alpha}, \alpha \in \{m, s\}$ and $(e_i)_{i=1,2,3}$ are the standard basis vectors in $\mathbb{R}^3$. $\mathbf{A} \in \mathbb{R}^{3N \times 3N}$ is usually referred to as the stiffness matrix and $\mathbf{f} \in \mathbb{R}^{3N}$ as the right-hand side in the FEM context.  The following subsections show, how the discretized system is solved by applying a change of basis on $\mathbf{A}$ which includes a mortar projection. A weak and discretized version $\mathbf{g}$ of $g$ is obtained as a byproduct of the computation of the mortar projection routine.

\subsection{Dual Mortar Approach} \label{sec:l2}
For the description of the dual mortar approach we follow \cite{Kra09, DK09_mb}. We  remind the reader, that in one of the earlier papers on the Mortar method \citep{BMP93}, the method was originally a domain decomposition method. Motivated by the need to solve partial differential equations on a domain with different non overlapping and non matching triangulations, the idea is essentially to "stitch" together the solution across the domain by enforcing a weak equality on the boundary of the two triangulations. Using our notation from the previous section, this implies $g=0$, and solving a linear elastic problem on the domain $\Omega = \Omega^m \cup \Omega^s$ under the condition
\begin{equation}
\int_{\Gamma_C^s}  [u] \, \tilde{\mu} \, d \gamma  = 0, \;  \tilde{\mu} \in \tilde{M}, \label{eq:mortar-cond}
\end{equation}
where $\tilde{M}$ is a modified trace space on the boundary of two different triangulations.

The dual mortar method we use here, differs from this original setting in three ways. First, we consider bodies which might initially not be in contact, i.e. $g \neq 0$. Second, in order to allow the bodies to separate from each other,  the equality condition in Eq.~\ref{eq:mortar-cond} becomes an inequality condition (see \cite{BHL98}).  And lastly, for the Lagrange multipliers we consider the convex cone $M^+$ defined as 
\begin{equation}
M^+ = \left \{ \mu \in M \big | \int_{\Gamma_C^s}  \mu  \, \lambda^s_+\, d \gamma \geq 0, \; \forall  \lambda_+^s \in X(\Gamma^s_C)^+ \right \} ,
\end{equation}
where $X(\Gamma^s_C)$ denotes trace space of $X(\Omega^s)$ on the boundary $\Gamma_C ^s$, $M$ the dual of $X(\Gamma^s_C)$ and $X(\Gamma^s_C)^+ :=\{ \lambda_+^s \in X(\Gamma^s_C), \, \lambda_+^s \cdot n^s \geq 0\}$. We can then replace the strong  non-penetration condition in Equation \ref{eq:nonpen} in such a way, that we only allow for penetration in the normal direction in a weak sense:
\begin{equation}
\int_{\Gamma_C^s} \big([u] -g  \big) \mu \cdot n^s\, d \gamma  \leq 0, \;  \mu \in M^+. \label{eq:weak-nonpen}
\end{equation}
In practice the weak non-penetration condition is difficult to enforce. Not only might the elements $u^s$ and $u^m$ initially be apart, even when the two bodies are in contact, i.e. $g=0$, the nodes on the surface triangulations are non matching. Thus, to relate the nodes of the master side to the slave side of the contact boundary, an additional operator is needed - our mortar projection $\Psi$.

The construction of $\Psi$ is then deduced in "reverse" by assuming the previously introduced mapping $\Phi$, which is part of the solution, already exists. To do so, we introduce the discrete trace spaces $X_h(\Gamma_C^m)$ and $X_h(\Gamma_C^s)$ of $X_h(\Omega^m)$ and $X_h(\Omega^s)$. For one coordinate they are spawned by the bases $(\lambda^m_i)_{i=1,...,N^m}$, $ (\lambda^s_i)_{i=1,...,N^s}$ with dimension $N^m$ and $N^s$.  $M_h^+$ denotes the set of the normal parts of the discrete variant of $M^+$and is spawned by biorthogonal basis functions $\mu_k$ with the same dimension as $X_h(\Gamma_C^s)$, i.e. $k=1,...,N^s$. The get their name from the biorthogonality condition they fulfill, which, by using $\delta_{ik}$ as the Kronecker symbol, reads:
\begin{equation}
\int_{\Gamma_C^s} \mu_k   \lambda^s_i \, d \gamma = \delta_{ik}   \int_{\Gamma_C^s} \lambda^s_i   \, d \gamma.
\label{eq:biorth_cond}
\end{equation}
We then demand that for the mortar projection $\Psi:  \Gamma_C^s \rightarrow \Gamma_C^s$ and the multiplier space $M_h$, the following weak equality holds for all elements $u_h^m \circ \Phi  \in X_h(\Gamma_C^s)$:
\begin{equation}
\int_{\Gamma_C^s} \big(\Psi (u_h^m \circ \Phi) - u_h^m \circ \Phi \big) \mu_h \;d\gamma = 0, \; \forall \mu_h \in M_h^+. \label{eq:mortarprojection}
\end{equation}
To get a discrete representation $\mathbf{T}$ of $\Psi$, we write the elements $u_h^m \circ \Phi$, $\Psi(u_h^m \circ \Phi)$ and $\mu_h$ in the basis representations  $u_h^m \circ \Phi = \sum_{i=1,...,N^m} v_i (\lambda^m_i \circ \Phi) , \, \Psi(u^m \circ \Phi) = \sum_{j=1,...,N^s} w_j \lambda^s_j$ , $\mu_h= \sum_{k=1,...,N^s} l_k \mu_k$, and reformulate Equation \ref{eq:mortarprojection} to:
\begin{equation}
 \sum_{i=1,...,N^m} v_i \int_{\Gamma_C^s}  (\lambda^m_i \circ \Phi)  \mu_k \, d \gamma = \sum_{j=1,...,N^s} w_j \int_{\Gamma_C^s}  \lambda^s_j  \mu_k \, d \gamma , \; \forall k = 1,...,N^s \label{eq:l2_discrete}.
\end{equation}
We now define  $\mathbf{D}:= ( d_{ki} )_{k,i = 1, ..., N^s}$ and $\mathbf{B}:= (b_{kj})_ {k=1,...,N^m, j=1,...,N^s}$ through:
\begin{align}
d_{ki} &:= \int\limits_{\Gamma^s_C} \lambda^s_i \mu_k  \, d \gamma \, \mathop{Id}   \quad \text{and} \\ 
b_{kj} &:=  \int\limits_{\Gamma^s_C} ( \lambda^m_j \circ \Phi) \, \mu_k  \, d \gamma \, \mathop{Id},
\end{align}
where $\mathop{Id} \in \mathbb{R}^{3x3}$is the identity operator.  We then reformulate Equation \ref{eq:l2_discrete} with $\mathbf{v}=(v_i)_{i=1,...,N^m}$ and $\mathbf{w}=(w_i)_{i=1,...,N^s}$ as
\begin{equation}
\mathbf{B}\mathbf{v} = \mathbf{D}\mathbf{w},
\end{equation}
and with $\mathbf{T}:= \mathbf{D}^{-1} \mathbf{B}$, we get the discrete representation of the mortar transfer operator.  Its purpose is to represent elements of $u_h^m$ of $X_h(\Gamma_C^m)$ via $\Phi$ in $X_h(\Gamma_C^s)$ in a meaningful way, using the weak equality in Equation \ref{eq:mortarprojection}. Computing the possibly dense inverse $\mathbf{D}^{-1}$ could potentially add additional computational complexity, but due to our specific choice of $M_h$ and condition Eq.\ref{eq:biorth_cond}, $\mathbf{D}$ becomes an easily invertible diagonal matrix. With $\mathbf{T}$ it is now possible to formulate the non-penetration condition (compare Eq. \ref{eq:nonpen}) in its discrete form at each node $p$ on the slave side $\Gamma^s_C$:
\begin{equation}
\big((u_h^s)_p - (\mathbf{T} u_h^m)_p\big)\cdot \mathbf{n}_p^s - g_p \leq 0, \quad p=1,...,N^s.  \label{eq:discrete-nonpen}
\end{equation}
Where $(u_h^s)_p$, $(u_h^m)_p \in \mathbb{R}^3$ denote local nodal vectors at node $p$, $\mathbf{n}_p^s$ the normal vector at node $p$, and $\mathbf{g}=(g_p)_{p =1,..., N^s}$ is the weighted gap vector defined as:
\begin{equation}
g_p := (D^{-1} g')_p, \quad (g')_p := \Big( \int_{\Gamma^s_C } \mu_p \, g \, d\gamma \Big) \, n^s_p , \quad p=1,...,N^s.
\end{equation}
Computing the entries  $d_{ki}$ and $b_{kj}$, which are surface integrals on a trace space, is nontrivial:  Apart from computing an approximation to $\Phi$, one needs to find suitable quadrature points on  $\text{supp}(\lambda_i^s)  \cap  \text{supp}(\lambda_i^m \circ \Phi) $, which is especially tedious in cluster computing when the meshes  $\mathcal{T}_h^s$ and $\mathcal{T}_h^m$ are  distributed across several compute nodes.  We use the library MOONoLith to obtain the quadrature points, as well as approximations to $\Phi$ , $n^{\Phi}$ and $g$, and we refer to \cite{DK09_mb} and\cite{KZ16} for an in-depth description of the required procedures, i.e. detection of intersections between the non matching meshes,  load distribution and parallel communication.

\subsubsection{Remark}
By defining the bilinear form $b: H_0^{-1/2}(\Gamma^s_C) \times X(\Gamma^s_C) \rightarrow \mathbb{R}, \quad b(\mu, v) := \langle \mu ,  [v]\rangle$, where  $\langle \cdot , \cdot \rangle$ denotes the dual pairing, the contact problem can be formulated as a saddle point problem. Find $(\mu, u) \in M_h^+ \times X_h$ such that: 
\begin{align}
a(u_h,v_h) +b(\mu_h, v_h) &=  f(v_h), \quad \forall \, v_h \in X_h , \\
b(\mu_h, u_h) &\leq  \langle \mu_h , g \rangle , \quad  \forall \, \mu_h \in M_h^+.
\end{align}
One can show that a uniform inf-sup condition holds and under additional mild assumptions the following a-priori error estimate between the discrete solution $(\mu_h, u_h)$ and the real solution $(\lambda, u)$ can be proven:
\begin{equation}
||u- u_h|| + ||\mu - \mu_h ||_{H^{-\frac{1}{2}}(\Gamma_C^s) } \leq C\, h^{\frac{1}{2} + \alpha} \sum_{k \in {s,m}} |u^k|_{\mathbf{H}^{\frac{3}{2}+\alpha}(\Omega^k)} ,\, \forall \alpha \in (0,\frac{1}{2}),
\end{equation}
with $u^k \in \mathbf{H}^{\frac{3}{2}+\alpha}(\Omega^k) = [H^{\frac{3}{2}+\alpha}(\Omega^k)]^3$  (from \cite{DK09_mb}, see also the references cited therein). 

\subsection{Algebraic formulation and solution}
We apply an orthogonal transformation $\mathbf{O}$ and a mortar transformation $\mathbf{Z}$ on the assembled discrete finite element system. The orthogonal transformation was introduced by \cite{KK01} and consists of Householder reflections at each contact node, whose purpose is to rotate the local coordinate systems at the nodes of the slave boundary $\Gamma_C^s$ in the direction of the local normals and align it with the constrained direction. The matrix $\mathbf{Z}$ algebraically transforms the two-body problem into a one-body problem (see remark 3.2 in \cite{Kra09}), thereby decoupling the non-penetration constraints $[u] \leq g$ (Eq.~\ref{eq:nonpen}). Consequently we use $\mathbf{u}$ for the coefficients of the solution vector in the standard basis and $\hat{\mathbf{u}}$ for the transformed basis. The solution is then be obtained from a quadratic optimization problem with $\hat{\mathbf{A}} := (\mathbf{O}\mathbf{Z})\,\mathbf{A}\,(\mathbf{O}\mathbf{Z})^t$ and $\hat{\mathbf{f}}:= (\mathbf{O}\mathbf{Z}) \, \mathbf{f}_h$: Find a $\hat{\mathbf{u}}$, satisfying Eq.~\ref{eq:discrete-nonpen}, such that
\begin{align}
\frac{1}{2} \hat{\mathbf{u}}^t \hat{\mathbf{A}} \hat{\mathbf{u}} - \hat{\mathbf{f}}^t \hat{\mathbf{u}} \quad \text{ is minimal}. \\ 
\label{eq:transformed_problem}
\end{align}

To construct $\mathbf{O}$ and $\mathbf{Z}$, we label the nodes of the discretized finite element system with three index sets $\mathcal{M}, \mathcal{S}$ and $\mathcal{I}$. The indices of the nodes on the boundary $\Gamma_C^m$ on the master side are in $\mathcal{M}$, those at the slave side in $\mathcal{S}$ and the indices of the remaining nodes in the interior are in $\mathcal{I}$.
For the orthogonal transformation $\mathbf{O}$, we then use outer normals $\mathbf{n}^s_i, \, i \in \mathcal{S}$ at nodes of the slave side of the contact boundary to compute local householder transformations. These transform the local coordinate systems on the slave side of the contact boundary in such a way, that the first coordinate of each nodal vector points in the direction of the local normals:
\begin{equation}
\mathbb{R}^{3N\times3N}   \ni \mathbf{O} := o_{ij} =\left\{
                \begin{array}{ll}
                  \mathop{\mathbf{id}}_{3x3} - \frac{2}{(\mathbf{n}^s_i)^t \mathbf{n}^s_i} \mathbf{n}^s_i (\mathbf{n}^s_i)^t  &, i=j \text{ and }  i  \in \mathcal{S}\\
                  \mathop{\mathbf{id}}_{3x3} &, i=j \text{ and }  i \in \mathcal{I} \cup \mathcal{M}\\
                  0 &,else .
                \end{array}
              \right.
 \label{eq:orthogonal_op}
\end{equation}
This has the effect that the constraints, introduced by the gap function $g$, only need to be considered on the first coordinate in each local coordinate system (see also\cite{Kra01,KK01,WK03}). Consequently, we assemble $\mathbf{g} \in \mathbf{R}^{3N}$ such, that we have the constraints $(g_i, \infty, \infty)^t$ for the coefficients $\hat{\mathbf{u}}_i \in \mathbb{R}^3$ on the slave nodes $i \in \mathcal{S}$, whereby $\infty$ signifies no constraint.

The second transformation $\mathbf{Z}$ uses the mortar operator $\mathbf{T}$ from the previous section and is formed in the following way:
\begin{equation}
\mathbf{Z} := \left[
\begin{array}{c c c}
\mathop{\mathbf{Id}}_{\mathcal{I}} & 0 & 0\\
0 & \mathop{\mathbf{Id}}_{\mathcal{M}} & \mathbf{T}^t\\
0 & 0 & \mathop{\mathbf{Id}}_{\mathcal{S}}\\
\end{array}
\right]  \in \mathbb{R}^{3N\times3N} .
\label{eq:mortar_op}
\end{equation}
The multiplication with $Z$ transfers the displacements of the master nodes to the slave nodes and transforms the two-body problem into a one-body problem.  To illustrate this, we apply the operator $\mathbf{Z}^{-t}$ to the solution $\mathbf{u}$:
\begin{equation}
\left[
\begin{array}{c c c}
\mathop{\mathbf{Id}}_{\mathcal{I}} & 0 & 0\\
0 & \mathop{\mathbf{Id}}_{\mathcal{M}} & 0\\
0 & -\mathbf{T} & \mathop{\mathbf{Id}}_{\mathcal{S}}\\
\end{array}
\right] 
\mathbf{u} = 
\left[
\begin{array}{c}
\mathbf{u}_{\mathcal{I}}\\
\mathbf{u}_{\mathcal{M}}\\
-\mathbf{T} \mathbf{u}_{\mathcal{M}} + \mathbf{u}_{\mathcal{S}}\\
\end{array}
\right] = \left[
\begin{array}{c}
\hat{\mathbf{u}}_{\mathcal{I}}\\
\hat{\mathbf{u}}_{\mathcal{M}}\\
\hat{\mathbf{u}}_{\mathcal{S}}\\
\end{array}
\right] =: \hat{\mathbf{u}} \quad \in \mathbb{R}^{3N} .
\end{equation}
The constraints of the transformed system are now only on the third component $\hat{\mathbf{u}}_{\mathcal{S}}$ of the transformed solution vector $\hat{\mathbf{u}}$ (compare also with Eq.~\ref{eq:discrete-nonpen}).  

The system can be solved  with semismooth Newton methods \cite{HIK02}, or, what is another appeal of this transformation, with monotone multigrid methods of optimal complexity \cite{Kra01,Kra09,DK09_mb}. Unfortunately Hertzian contact theory makes several assumptions that are not necessarily met when dealing with real complex surfaces:  small strains, elastic material, absence of  friction, a relatively small contact area, each body  being in essence a half space and lastly, small strains, close proximity and continuous surfaces.  To  accommodate  for  this, we apply a non-smooth SQP method similar to \citep{KM11} in which we add an outer loop to the solution process.  The contact problem is then solved as a sequence of $k=1,..., n_{\text{outer}}$ contact problems, where in each iteration we adjust for the  new obstacle configuration, i.e. we recompute the gap, the normal field and the mortar operator (see Alg. \ref{alg:sqp} where we added the subscript $k$ for all iteration dependent terms).

\begin{algorithm}
\begin{algorithmic}
\STATE Initialize $k$, $\hat{\mathbf{u}}_0, \mathbf{O}_0,\mathbf{Z}_0, \mathbf{g}_{0},  \mathbf{A}, \mathbf{f}_0$
\FOR{$k=0$ to $n_{\text{outer}}$}
 \STATE Solve $\hat{\mathbf{u}}_{k+1} = \underset{\hat{\mathbf{u}}}{\mathrm{argmin}} \frac{1}{2} \hat{\mathbf{u}}^t \hat{\mathbf{A}}_k \hat{\mathbf{u}} - \hat{\mathbf{f}}_k^t \hat{\mathbf{u}} \; \text{with:} \; \hat{\mathbf{u}} \leq \mathbf{g}_{k}$
 \STATE Move meshes by applying displacement $\mathbf{u}_{k+1} - \mathbf{u}_{k}$: $\mathcal{T}^{\alpha}_{k} \mapsto \mathcal{T}^{\alpha}_{k+1}, \alpha \in \{m, s\}$.
 \STATE Recompute $\mathbf{O}_{k+1}$, $\mathbf{Z}_{k+1}$,$\mathbf{n}_{k+1}$,and $\mathbf{g}_{ k+1}$ for step $k+1$.
\ENDFOR 
\end{algorithmic}
\caption{Non-smooth SQP method.}
\label{alg:sqp}
\end{algorithm}

Note that we have chosen a fixed $n_{\text{outer}}$ as stopping criterion, which was heuristically determined from solving the benchmark problem in section \ref{sec_res_hertz}. Alternatively, one could also define a relative criteria such the norm of the step length, etc.
\subsection{Implementation}
\begin{figure}[hbt]
\begin{center}
\includegraphics[width=0.8\textwidth]{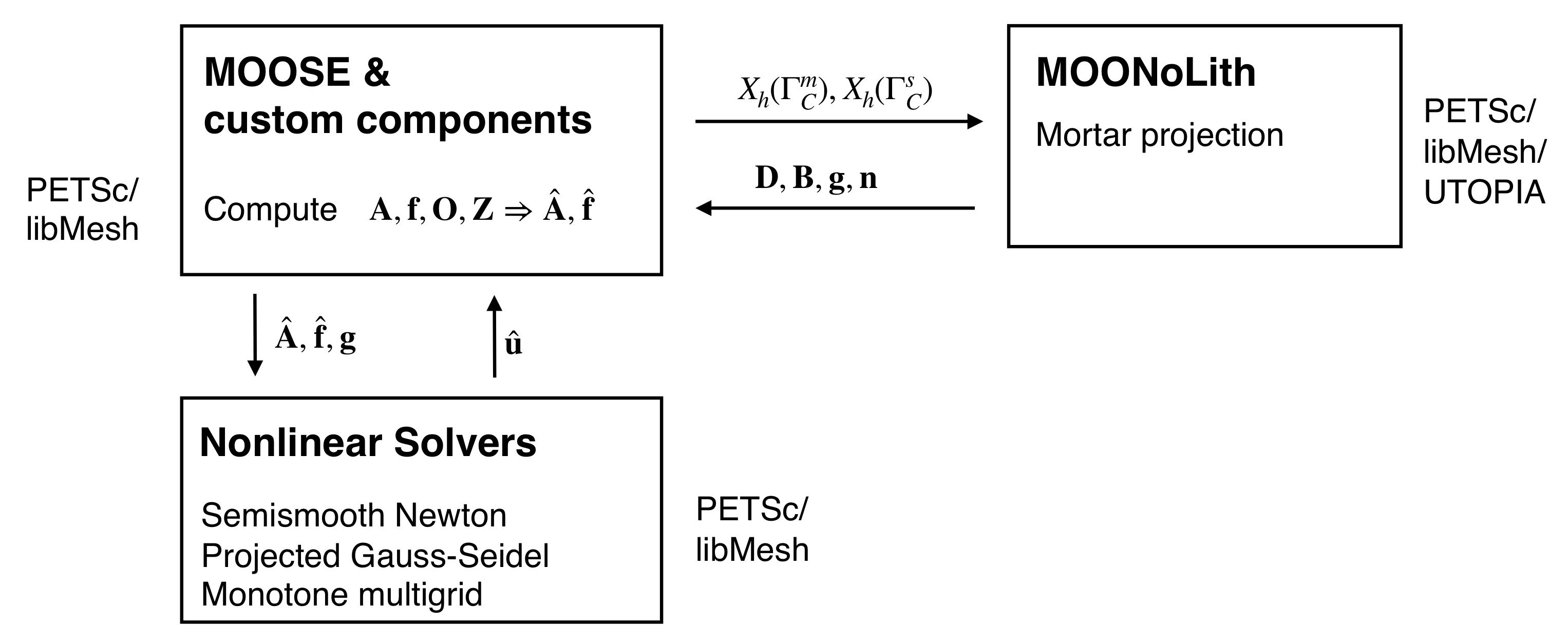}
\caption{}
\label{fig:schematic_framework}
\end{center}
\caption{Schematic of the software components and their interaction.}
\end{figure}

Our implementation is based on the MOOSE framework \cite{GNH+09} and MOONoLith \cite{KZ16}. In geophysical applications, MOONoLith is primarily used to compute transfer operators between different, nonmatching triangulations (see for example \cite{PVC+19_a,PVC+19_b, schaedle_2019}). Here it is used to compute the previously described matrices $\mathbf{D}$ and $\mathbf{B}$ of the mortar projection, as well as the gap  $\mathbf{g}$ and the local normals $\mathbf{n}$. All matrices and vectors are integrated in MOOSE using several custom made MOOSE components, in which we also carry out the transformations described in the previous sections. To solve variational inequalities, such as contact problems, we have augmented the MOOSE framework with a new executioner class for the non-smooth SQP method and added a set of dedicated solvers. These include semismooth Newton, projected (nonlinear) node-based Gauss-Seidel, and monotone multigrid, all of which are implemented as SNES solvers in PETSc \cite{BAA+17}. The overview of the framework is shown in Fig.\ref{fig:schematic_framework}, whereby the operators refer to Eqs.\ref{eq:discrete_operator},\ref{eq:transformed_problem}, \ref{eq:orthogonal_op}, \ref{eq:mortar_op}.  We have also included libMesh \cite{KPH+06}, PETSc and UTOPIA \cite{ZKN+16} in the figure, which were used to create the interfaces. The integration of oure components into MOOSE can be considered deep in the sense, that while the interfaces were designed as MOOSE components, calls to libMesh objects and methods within MOOSE, as well as to PETSC structs and functions, were used throughout the implementation. We like to remark that these libraries were chosen in particular, because they are proven, suitable for parallel cluster computing and open source. 
\section{Numerical experiments} \label{sec:results}
\subsection{Hertzian contact}
\label{sec_res_hertz}
\begin{figure}[hbt!]
\centering
    \begin{subfigure}[b]{0.45\textwidth}
        \includegraphics[width=0.9\textwidth]{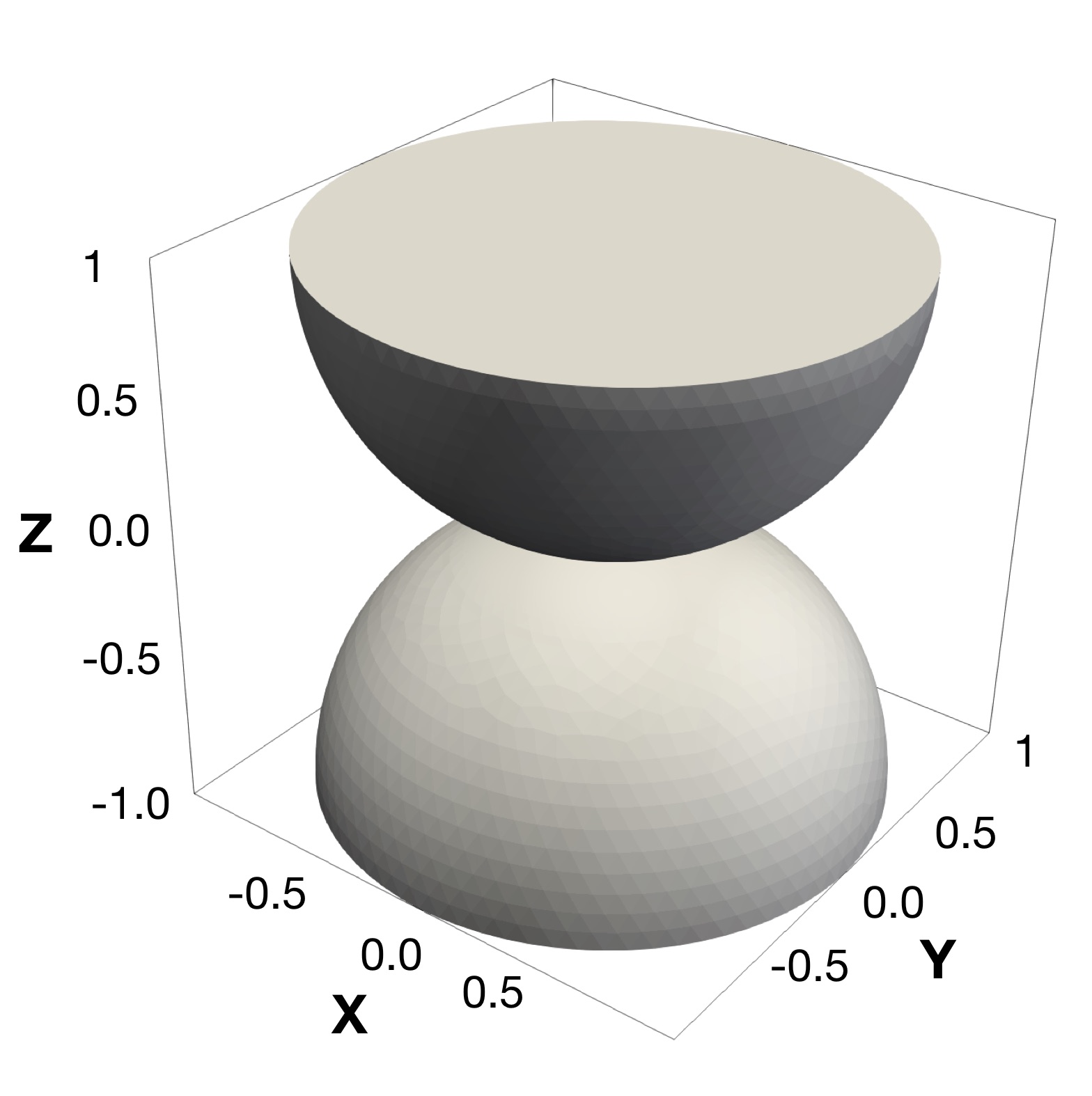}
        \caption{Setup of the Hertzian contact problem, showing two half spheres in contact.}
        \label{fig:hertz_setup}
    \end{subfigure}
    \hspace{0.2cm} 
    \begin{subfigure}[b]{0.5\textwidth}
        \includegraphics[width=0.9\textwidth]{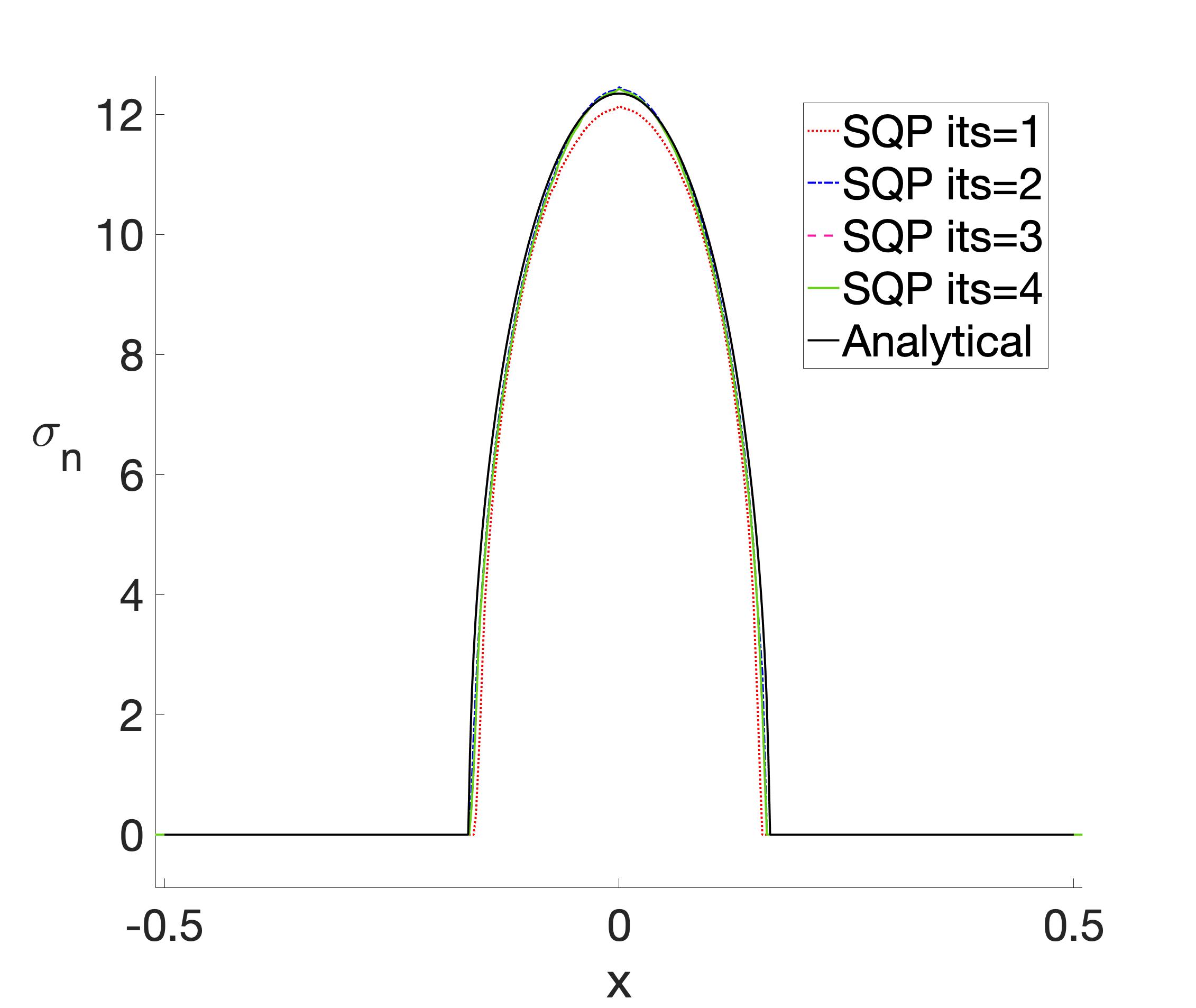}
        \caption{Resulting contact stress distributions for several SQP iterations compared to analytical solution.}
        \label{fig:hertz_bnd_stresses}
    \end{subfigure}
    \caption{Hertzian Contact}
    \label{fig:hertz_exp}
\end{figure}

To validate the method and to determine a reasonable number of SQP iterations, we set up a Hertzian contact problem. The term stems from a study by Heinrich Hertz in which he derived analytical solutions for two bodies in contact, which have elliptic contact interfaces \cite{Her1882}. In particular, in these kind of problems, one seeks to replicate the parabolic shape of the boundary stresses in the contact region.
Here, we set up a two-body problem in which two half spheres are in contact (Fig.~\ref{fig:hertz_setup}). Both half spheres have radius $1$, Youngs Modulus $E=100$, Poisson ratio $\nu=0.3$, and are moved into each other by a distance $d=0.05$. The meshes of the two bodies have 330'379 nodes and are refined at the contact zones with a mesh width of roughly $5 \times \, 10^{-5}$.  We then solve the Hertzian contact problem according to algorithm \ref{alg:sqp} with 1-4 SQP iterations. To compare the results with the analytical solution, we infer the contact radius from the numerical result with 4 iterations and compute the boundary stresses analytically. The results are shown in Fig.~\ref{fig:hertz_bnd_stresses}:  The analytical solution is depicted as a solid black line. The solution after one SQP-iteration (dotted red line) deviates visibly from the analytical solution. After the second SQP iteration (blue line), the solution is almost congruent with the analytical solution and further improves slightly after three (dashed purple line) and four (solid green line) SQP iterations. As a consequence, we restricted ourselves to the use of three SQP iterations in the subsequent numerical experiments.

\begin{figure}[bt]
\begin{center}
{\small
\includegraphics[width=0.95 \textwidth]{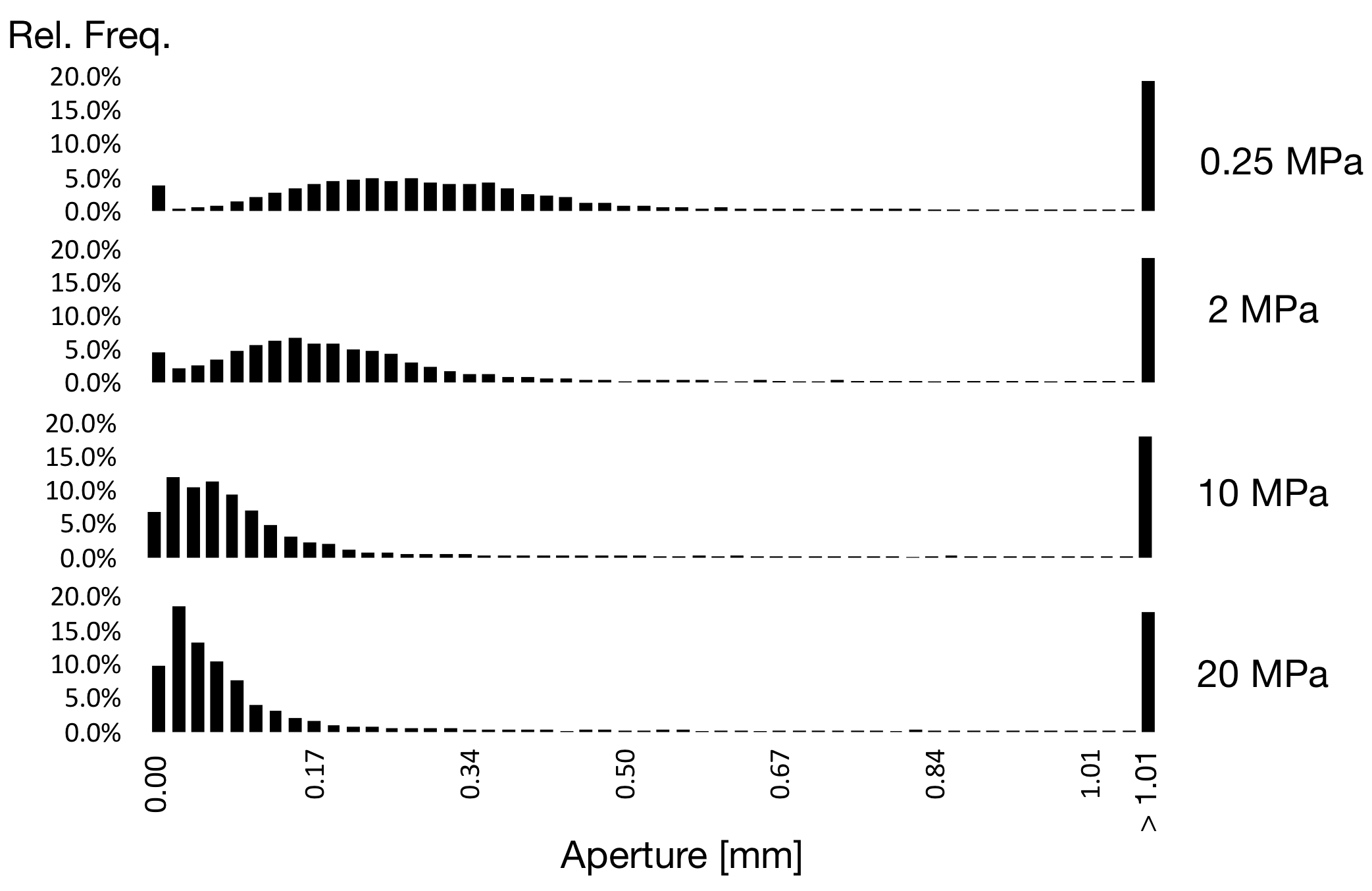}
}
\caption{Histograms of the aperture distribution at increasing confining stresses of 0.25, 2, 10 and 20~MPa. As the confining stress increases, areas with large apertures become less frequent, while the frequency of small-aperture regions increases. }
\label{fig:aperture_hist}
\end{center}
\end{figure}
\subsection{Contact between two rough rock surfaces}
The capability of the presented methodology to solve contact problems between highly complex surface topographies is demonstrated in a numerical experiment with a rock specimen taken from the Grimsel underground rock laboratory in Switzerland. The rock fracture geometries are adapted from previous studies \citep{vogler_2016b, vogler_2016c, vogler_2018} and are embedded in a cylindrical specimen, consisting of granodiorite rock (Fig. \ref{fig:mesh}). Here, an increasing compressive load is applied to the specimen cylinder top in the axial direction (z-direction), while displacement in the z-direction is fixed to zero, with a Dirichlet boundary condition on the cylinder bottom.  The diameter of the cylinder is 122mm and the material is defined to be linear elastic with a Youngs Modulus of $E=10 \text{MPa}$ and a Poisson ratio of $\nu=0.33$.
Fracture surfaces were digitized with photogrammetric techniques \citep{vogler_2017,vogler_2017b,vogler_2018}, which produce a mesh of triangular elements at the body surface. From the surface mesh, a volumetric mesh of tetrahedral elements can be generated for the solid bodies with the software TRELIS \citep{trelis}.  The upper body contains 101'637 nodes and the lower body contains 98'866 nodes, which results in 601'509 degrees of freedom for the simulation.  The upper contact boundary consists of 8793 nodes, the lower one has 7898 nodes. Since the meshes are non-uniform, these numbers are difficult to relate to a resolution, which is by definition a measure that only applies to uniform meshes. We would argue, however, that non-uniformness gives us at least the same, or even a higher, "effective" resolution as a uniform mesh, since the nodes are distributed more effectively in our case, as they are distributed according to the complexity of the rock surface (see Figure \ref{fig:mesh}). The simulations themselves ran on 4 nodes  (2~x Intel Xeon E5-2650~v3 @~2.30GHz) with 10 CPU's each, on the cluster of the  Institute of Computational Science in Lugano, Switzerland.

\begin{figure}[htbp]
\begin{center}
{\small
\includegraphics[width=0.6\textwidth]{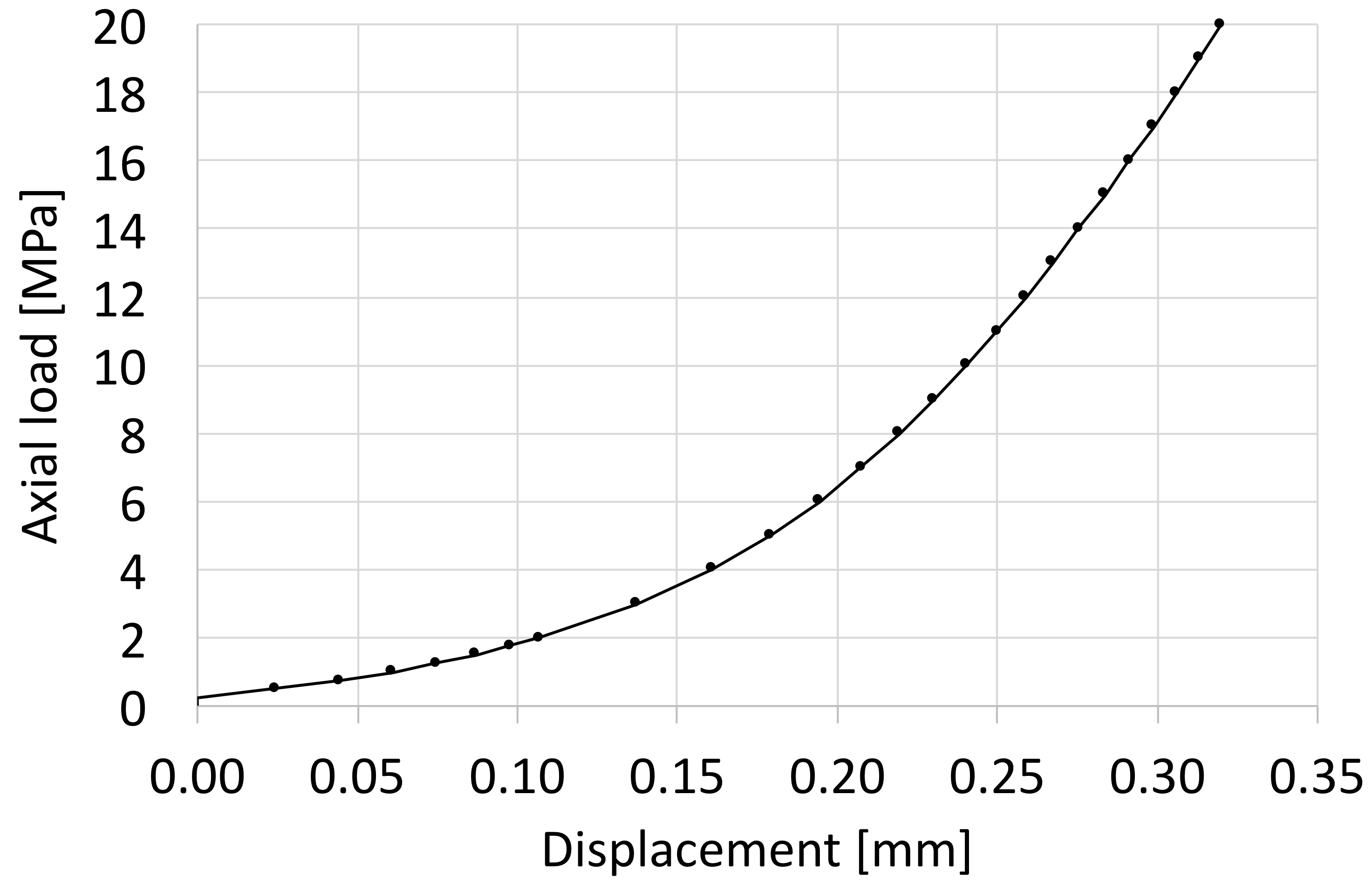}
}
\caption{ Simulated loading curve, showing displacement (i.e., fracture closure) versus axial load.}
\label{fig:loading_curve}
\end{center}
\end{figure}

In this section, aperture is defined as the values of the gap function $g$, which reside on the nodes of the slave side.  In its implementation, $g$ differs from theory, as on a finite element mesh, normals are defined on the surfaces of the element sides and not at the nodes. The direction of $g$ is therefore computed as the average of the normals of all sides of the mesh surrounding each node of the contact side. Still, we believe this to be a more accurate description of the distance between the fracture surfaces than a simple projection in the normal $(0,0,-1)^t$-direction and we thus use the value of $g$ as aperture. For the assembly of the mortar operator, only intersections up to a reasonable distance of 0.21~mm are detected and the maximum value of the aperture is fixed at 0.21~mm. This does not impact the accuracy of the contact method, as the displacements of the rock are smaller than this maximum value (compare also with Figure~\ref{fig:loading_curve}).

\begin{figure}[bt]
\begin{center}
\includegraphics[width=1.0\textwidth]{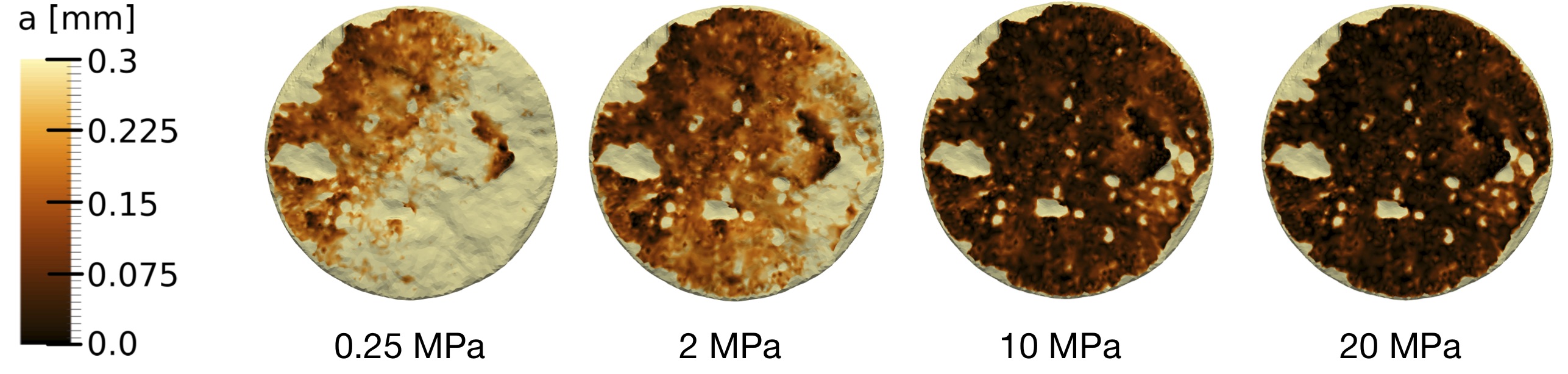}
\caption{View of the different aperture fields under increasing confining stresses of 0.25, 2, 10 and 20~MPa.}
\label{fig:apertures}
\end{center}
\end{figure}

\begin{figure}[bt]
\begin{center}
\includegraphics[width=1.0\textwidth]{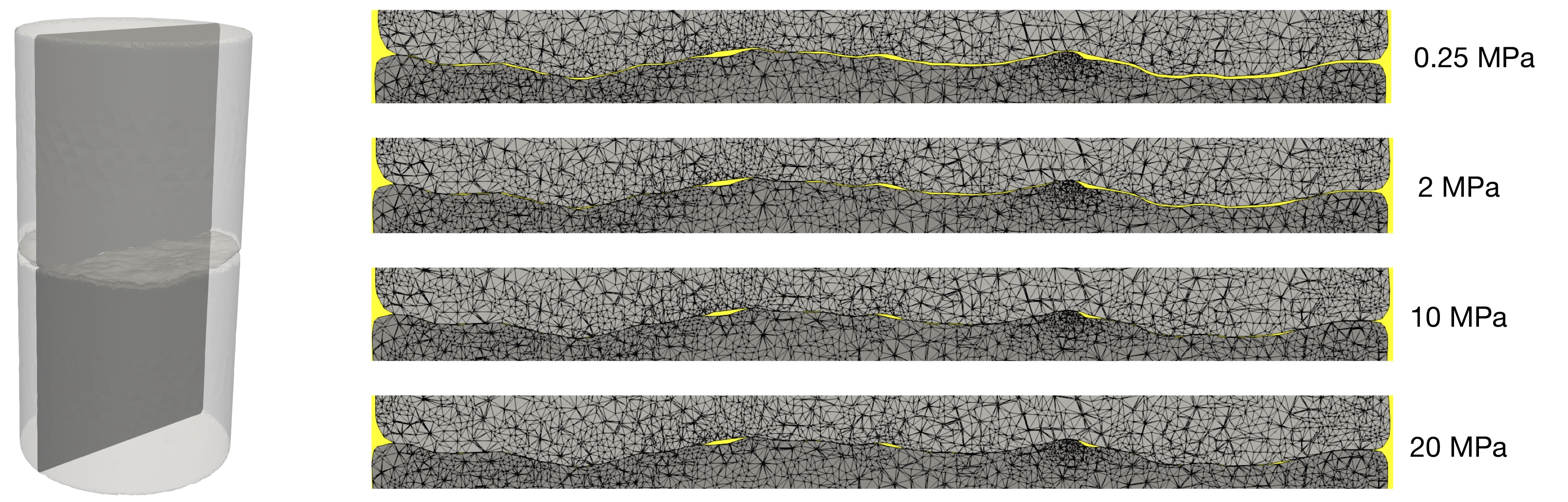}
\caption{Left: 3D representation of the approximately horizontal fracture. Right: 2D view of the fracture, shown on the left. Open regions of the fracture are shown in yellow. }
\label{fig:closure_clip}
\end{center}
\end{figure}

Figure~\ref{fig:apertures} shows the closure of the fracture aperture field under increasing normal loads, from 0.25 to 20~MPa. The aperture field is highly heterogeneous across the interface, with isolated regions of small apertures, for example, right of the center of the surface. Increasing the confining stress transforms the field significantly: At 0.25~MPa, only few parts of the surface are in contact and the apertures across large parts of the surface are at the maximum value of 0.21~mm. The aperture field then decreases significantly for a confining stress of 2~MPa and even more so at 10~MPa. There are regions, however, which do not close, so that apertures of at least 0.21~mm remain, even when the confining stress is further increased to 20~MPa. The closing of the fracture is again illustrated in Figure~\ref{fig:closure_clip}, where we show a cross section of the fracture during closure with the open part of the fracture depicted in yellow. At 0.25~MPa, there are only few contact areas and a large part of the fracture is still open. One can observe how the fracture closes more and more with increasing normal stress until, at 20~MPa, the only open areas are essentially those, where the fracture geometry indicates  cavities. In Figure~\ref{fig:aperture_hist}, we illustrate the overall behavior of the aperture field during closure. As the confining stress increases, the histogram is shifted to the left and the aperture field is distributed around a lower mean. 

Hence, our method replicates the closure of the fracture in contact, free of over-closure. Studying the aperture field, the cross-section and the aperture histogram (Figure~\ref{fig:apertures}, \ref{fig:closure_clip}, \ref{fig:aperture_hist}), we see that even at confining stresses of 10 and 20~MPa, the fracture is far from closed. Small cavities and channels still exist and the deformed fracture geometries can be used for simulating fluid flow under increasing confining stresses.
\begin{figure}[bt]
\begin{center}
{\small 
\includegraphics[width=1.0\textwidth]{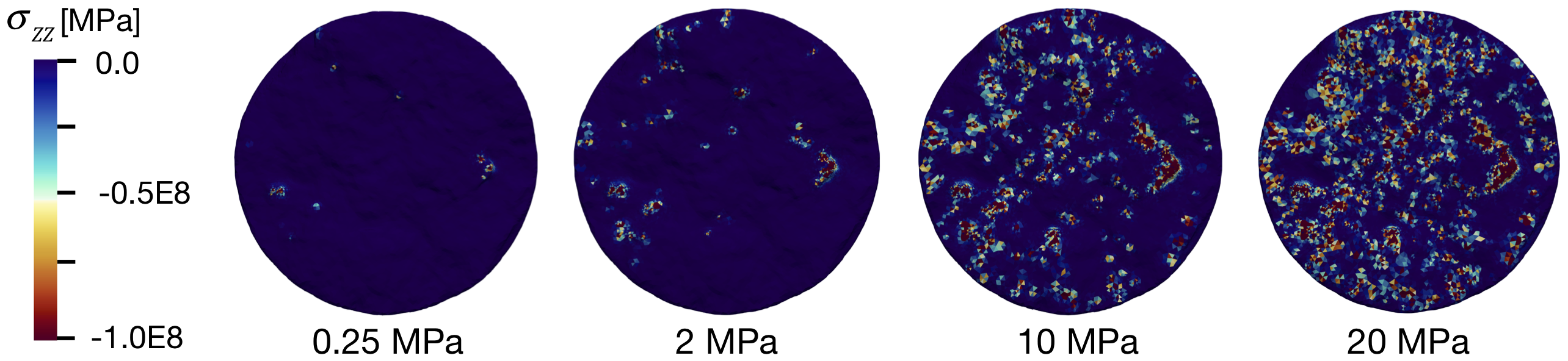}
}\\[5pt]
{\large \hspace{1.8cm} Stresses in normal direction $\sigma_{zz}$}
\caption{Top view of the lower fracture surface for confining stresses of 2, 10 and 20~MPa. The color indicates the vertical stresses $\sigma_{zz}$.}
\label{fig:sigma_n}
\end{center}
\end{figure}
\begin{figure}[htbp]
\begin{center}
{\small
 \includegraphics[width=1.0\textwidth]{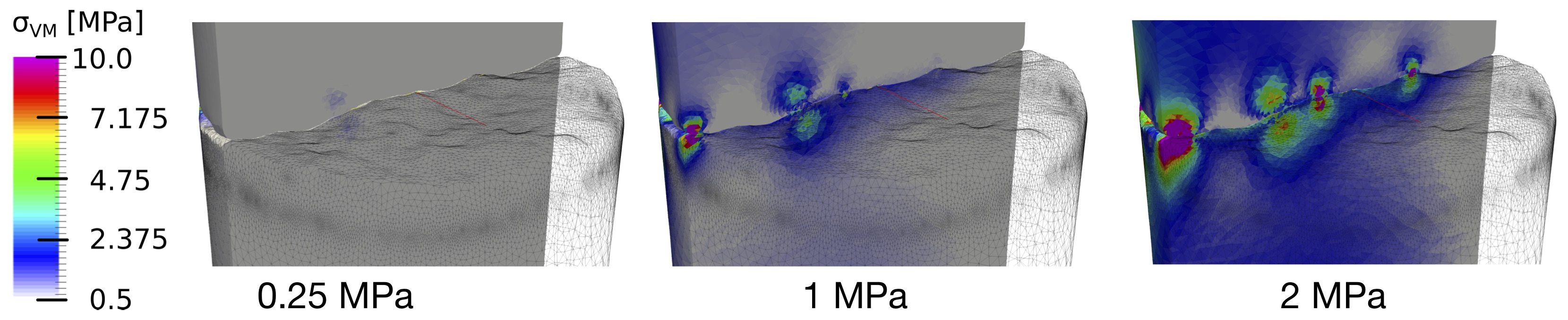}\\
}
\caption{The color indicates von Mises stresses around contact points during the closure of the fracture at confining stresses of 0.25, 1 and 2~MPa.}
\label{fig:vonmises}
\end{center}
\end{figure}
The development of the vertical stresses is shown in Figure~\ref{fig:sigma_n}. The vertical stresses develop around the few existing contact points for an axial load of 0.25~MPa and spread over the contact surface when  loads of more than 10~MPa are applied. Comparing Figures~\ref{fig:apertures} and~\ref{fig:sigma_n}, we observe, that the first vertical stresses form around regions with small apertures (dark parts), which become more pronounced when the confining stresses are increased. In Figure~\ref{fig:vonmises}, we show a cross section of the fracture with the von Mises stresses, developing at the contact nodes under increasing loads. In contrast to simpler contact models, we are able to observe, that the stresses develop from the contact surface in non-orthogonal directions in the interior of the body. 
This enables the observation of stress concentrations and stress shadows around contact regions.

While this study focuses primarily on contact detection, normal stresses, aperture field development and fracture closure, detailed knowledge of stress field variations around contact zones is of imminent importance. 
This is the case, as localized stress variations around zones of contact can lead to plastic deformation and failure, which permanently alter the mechanical and hydraulic behavior of the fracture \cite{vogler_2016}. 
The extent of stress concentrations around contact zones depends on both the external load as well as on fracture surface topographies. The presented approach therefore enables an estimation of stress extremes for fracture topographies commonly encountered in some specific rock types. 

Rock fractures subjected to normal loading show a characteristic nonlinear fracture closure curve, where fracture closure becomes increasingly smaller for the same load increment, until it approaches the behavior of elastic deformation in a solid body \cite{bandis_1983,matsuki_2008,zangerl_2008,vogler_2016,vogler_2018,kling_2018}.  The fracture closure curve, obtained from our numerical experiment, is shown in Figure~\ref{fig:loading_curve} for increasing axial loads from 0.25 to 20~MPa.  Here, we increased the axial load incrementally in steps of 0.25~MPa from 0~to 2~MPa and then in steps of 1~MPa from 2~to 20~MPa.  The fracture closure (displacement) is measured as the average displacement of nodes in zones of about 1~mm thickness, approximately 2.5~cm above and 2.5~cm below the fracture.  Displacement is measured at such a small distance from the fracture to avoid large heterogeneities in displacement due to the heterogeneous distribution of surface height and contact area over the entire fracture surface. To isolate displacements of the fracture, elastic displacements caused by the deformation in the lower specimen half, are excluded by subtracting the average displacements 2.5cm below the fracture from the average displacements 2.5cm above the fracture(see also \cite{vogler_2018}).
The curve in Figure~\ref{fig:loading_curve} shows, that the displacements become smaller when the confining stresses become larger, eventually leading to a linear relationship between axial load and displacement. The shape of the curve can be explained by noting that an increase in loading  results in more areas of the fracture being in contact, increasing the overall resistance to the confining stress until a quasi-linear elastic response is reached. This is a well-known characteristics of loading curves and, together with the obtained boundary stresses for the Hertzian contact, underlines the soundness of our approach.
\section{Conclusions}
We implemented a dual mortar approach to compute the contact between two rough fracture surfaces with non-matching meshes, employing linear elasticity and linearized contact conditions. Unlike penalty methods, neither solution nor convergence depend on an external parameter and over-closure of the fracture is not a concern. 
To test our approach, we used complex fracture geometries, obtained from an actual granite rock. The high-resolution 3D fracture geometries were resolved from a rock specimen that had undergone laboratory experiments. We have demonstrated the validity of our method by  reproducing the boundary stresses for Hertzian contact and the characteristic nonlinear closing behavior of a fracture under increasing normal loads.  The presented methodology enables investigation of the stress field development and its variations in solid bodies, fracture aperture fields, contact areas and other behavior for arbitrary complex surface geometries. 

Our implementation uses open-source software components that are designed for parallel computing. In particular, we use MOOSE for the finite element assembly and MOONolith for the computation of the mortar transfer operator. Together with our contact formulation, we can extend our framework to three directions:  First, extend our formulation of frictionless contact to include friction as in \cite{Kra09,PWG+12}. Second, use our formulation as a stepping stone for a wider class of more efficient multigrid obstacle solvers \cite{Kra09}. And third, leveraging our implementation in MOOSE, to simulate fluid flow with deforming fractures as a fluid-solid interaction approach, which we have outlined in \cite{PVN+18}.

\section*{Acknowledgements}
We gratefully acknowledge funding by the Swiss Competence Center for Energy Research - Supply of Electricity (SCCER-SoE), by the Werner Siemens-Stiftung (Werner Siemens Foundation) and by ETH~Zurich. 

\section*{ORCID}
C. von Planta - https://orcid.org/0000-0002-3111-7186 \\
D. Vogler - https://orcid.org/0000-0002-0974-9240 \\
P. Zulian - https://orcid.org/0000-0002-5822-3288 \\
M.O. Saar - https://orcid.org/0000-0002-4869-6452 \\
R.H. Krause - https://orcid.org/0000-0001-5408-5271

\bibliography{bibliography}

\end{document}